\documentclass[aps,preprint]{revtex4-1}

\usepackage[dvipdf,]{graphicx}
\usepackage{graphicx}
\usepackage{float}
\usepackage{color}
\usepackage[caption=false]{subfig}
\usepackage{amsmath}
\usepackage{siunitx}
\usepackage{hyperref}

\newcommand{\me}{\mathrm{e}}
\newcommand{\mi}{\mathrm{i}}

\renewcommand{\tt}[1]{\mathrm{#1}}
\renewcommand{\vec}[1]{\mathbf{#1}}

\newcommand{\eq}{\begin{equation}}
\newcommand{\qe}{\end{equation}}

\renewcommand{\k}{\vec{k}}

\newcommand{\Q}{\vec{Q}}

\newcommand{\rr}{\vec{r}}

\newlength{\figwidth}
\newlength{\widefigwidth}
\begin{document}
\title{Validation of x-ray line-profile analysis of extended defects in deformed crystals}
\author{C P Race}
\email[Contact: ]{christopher.race@manchester.ac.uk}
\author{T Ungar}
\affiliation{Department of Materials, University of Manchester, Manchester, M13 9PL, United Kingdom}
\author{G Rib\'arik}
\affiliation{Department of Materials Physics, E\"otv\"os Lor\'and University Budapest, H-1117 P\'azm\'any P. s\'et\'any 1/A, Hungary}

\begin{abstract}
Quantitative measurements of extended defects in crystalline materials are important in understanding material behaviour. X-ray line profile analysis provides a complement to direct counting in the electron microscope, but is an indirect method and requires validation. Previous studies have focused on comparing x-ray analysis to electron microscopy results. Instead, we use simulated defective material with known defect content and apply line profile analysis to calculated diffraction profiles to directly show that line profile analysis can reliably quantify dislocations and stacking faults.

\end{abstract}

\maketitle


Extended defects in crystalline materials, such as dislocations and stacking faults, strongly influence material properties and play a key role in many failure mechanisms. To properly understand material behaviour, we therefore need reliable ways of quantifying and characterising these defects. One common approach is to directly image defects in the transmission electron microscope and count them. Though highly successful, this approach has several disadvantages: it is time consuming and expensive to prepare the thin foils needed for analysis; this preparation can strongly affect the defects that we wish to measure; not all the defects will be large enough to see; high defect densities make individual defects difficult to identify; and the volume of material analysed in a typical foil is only very small. 

An alternative approach is to use x-ray \emph{line profile analysis}. In this approach, we examine the one-dimensional profile of diffracted intensity as a function of angle, arrived at via an integration around the Debye-Scherrer rings formed by diffraction of a highly mono-chromated beam of x-rays. Such profiles are commonly generated from powder or polycrystalline samples, but with a bright enough source of x-rays (e.g. from a synchrotron) profiles from individual grains can be measured. The \emph{positions} of the peaks in the line profiles correspond to the \emph{spacing} of sets of planes in the sample and so can be used to determine crystal structure and macro strains. Line profile analysis takes this a step further by making use of the fact that the \emph{shapes} of the peaks are affected by the \emph{distortion} of the corresponding set of planes and so contain information about the types and numbers of defects present.

Because line profile analysis requires only simple sample preparation and can be used to quickly analyse a relatively large volume of material, it provides an important complement to direct counting of defects in the electron microscope. We need to remember, however, that it is an indirect method of quantifying defect content. It begins with physical models of the distortions created by individual defects and considers the aggregate effect that a population of such defects will have on the shapes of the peaks in a line profile. This model line profile is thus a function of parameters representing the nature, density and arrangement of defects within the material and the technique proceeds by optimising the values of these parameters to achieve a model profile that best fits the experimental profile. We are thus attempting to infer details of complex features of the sample from a relatively simple one-dimensional signal and the task would be hopeless, were it not the case that the response of the peak shapes varies from peak to peak, with diffraction order and with distance from the peak centre, in different ways for different defects and for different types and arrangements of dislocations.

The physical basis for line profile analysis is sound. The distortion due to a dislocation, for example, is well described by theory at various levels of physics, including for the case of a general burgers vector and line direction, and for anisotropic elasticity \cite{Bacon:1980fk}. However, given the complexity of the process, an empirical validation is required. Previous attempts to validate the results of line profile analysis have relied on comparisons with counts in the electron microscope \cite{Ungar:1984aa}, but this is a somewhat indirect approach: the electron microscope does not give the full picture. Here we undertake an alternative, more direct, validation.

Rather than consider an experimental sample of material, we instead take a sample of simulated material to which we can apply line profile analysis and for which we know the precise defect content. We proceed by taking an atomistic simulation cell containing $\sim 4$ million atoms in a face-centred cubic arrangement. We use an embedded atom method model for copper \cite{Ackland:1987hs} and randomly add six $1/2\langle 110 \rangle$-type dislocation loops of $\SI{12}{\angstrom}$ radius to act as ``seeds'' for plastic deformation. Using the classical molecular dynamics software Lammps \cite{Plimpton:1995fv} we briefly anneal the simulation cell for  $10\,\tt{ps}$ at $300\,\tt{K}$ then apply a uniaxial distortion at a constant engineering strain rate (again at $300\,\tt{K}$ and zero pressure) to a target overall strain. We then return the simulation cell to an approximately cubic shape via a biaxial strain in the other two directions. Both stages of the compression last $0.5\,\tt{ns}$. By varying the target strain (5\%, 10\%, 20\%, 40\%, 60\% and 80\%) we arrive at samples of material with widely varying defect content with dislocation densities between zero and $40\times 10^{16}\,\tt{m}^{-2}$ and stacking fault content from zero to 8\%, calculated using the Ovito software~\cite{0965-0393-18-1-015012}. To further vary the density and character of the defects we undertake brief annealing of the deformed samples for $0.5\,\tt{ns}$ at a temperature of $800\,\tt{K}$. Even this short anneal is long enough for the early stage relaxation of the dislocation distributions to take place~\cite{Race:2019ac}. Further details of our simulation approach are given in the companion paper \cite{Race:ts}.

\begin{figure}\begin{center}
{\includegraphics[width=\figwidth]{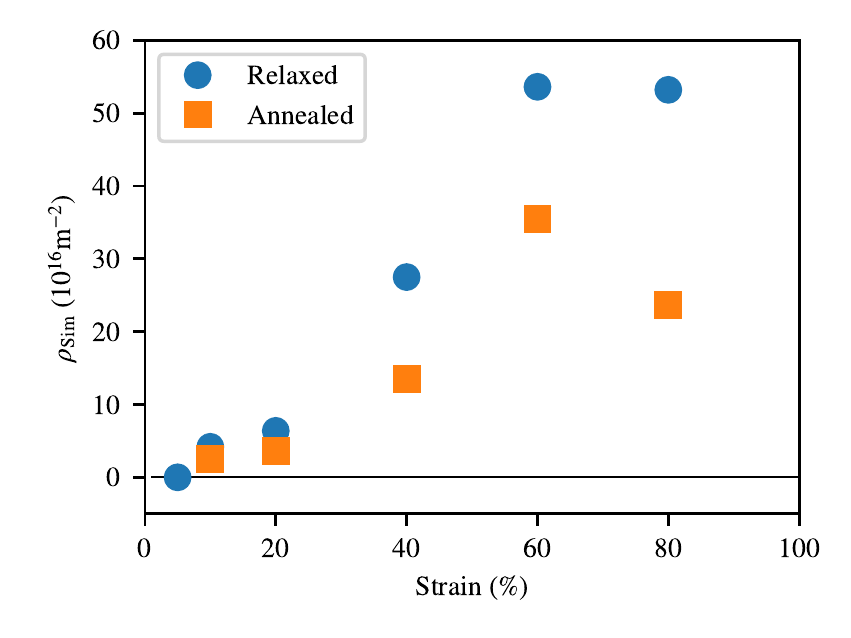}}
\caption{Dislocation content of the cold-worked cells as a function of strain for the as-worked (relaxed) and annealed ($0.5\,\tt{ns}$) cases.}
\label{fig:coldworked-simdensity}
\end{center}
\end{figure}

The next step is to calculate a diffraction line profile from each of the atomistic simulation cells. For an incoming x-ray beam with wavevector $\k$ scattered into an outgoing beam with wavevector $\k'$ by an array of atoms with positions $\{ \rr_i\}$ indexed by $i$, the relative phase difference for the scattered wave from the $i^{\tt{th}}$ atom will be $\exp(-\mi \Q\cdot\rr_i)$, where $\Q=\k'-\k$ and $\hbar\Q$ is the momentum transfer. Omitting a prefactor, which affects only the normalisation, the amplitude of the scattered wave in the direction of $\k'$ (assuming a fixed $\k$) is
\eq\label{eq:scattered_amplitude}
\Psi(\Q) =\sum_i \me^{-\mi  \Q\cdot \rr_i},
\qe
assuming that all the atoms in our sample are of the same type and the angular variation in scattered amplitude can be neglected. The scattered intensity is given by $I(\Q)=\Psi^{\ast}(\Q)\Psi(\Q) = \sum_{i,j} \me^{-\mi  \Q\cdot \rr_{ij}}$, where $\rr_{ij}=\rr_j-\rr_i$, and to arrive at the equivalent of a powder pattern (or pattern for an untextured polycrystal) we integrate this intensity over the full solid angle in $\vec{Q}$ assuming, without loss of generality, that for each atom pair $\rr_{ij}$ lies along $\theta=0$.

\begin{figure}\begin{center}
{\includegraphics[width=\figwidth]{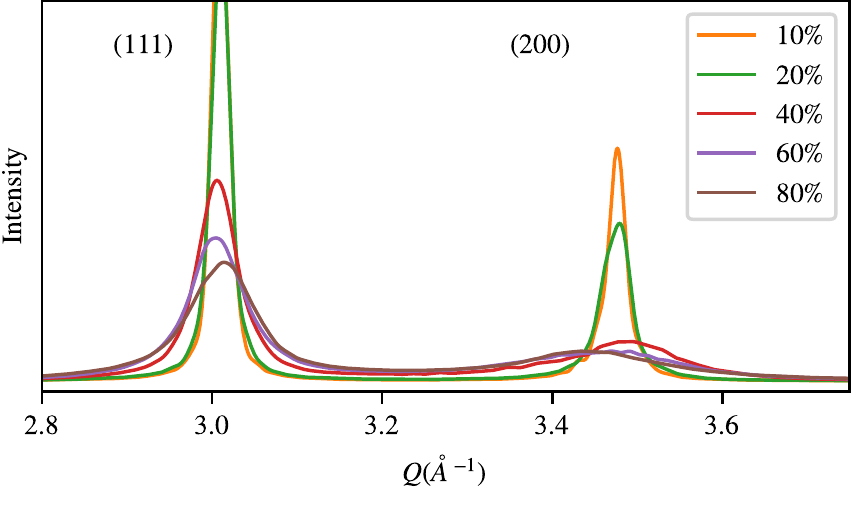}}\hfill
\caption{Examples of line profiles for two low-index reflections in the profiles calculated for the relaxed cold-worked material for a range of strains.}
\label{fig:lineprofiles}
\end{center}
\end{figure}

This leads to the \emph{Debye equation} \cite{Debye:1915aa}
\eq\label{eq:debye}
I(Q) = 4\pi \sum_{i,j}\frac{\sin (Q\,r_{ij})}{Q\,r_{ij}},
\qe
which allows us to calculate the whole line profile $I(Q)$ from a set of atomic coordinates. Because $I(Q)$ depends on $\{\vec{r}_i\}$ only through the interatomic separations $r_{ij}=|\rr_{ij}|$, we can implement~\eqref{eq:debye} by calculating a histogram of the interatomic separations with bin centres and frequencies $\{(R_s, f_s) \}$ so that the diffraction profile is given by
\eq
I(Q) = 4\pi \sum_{s}f_s\,\frac{\sin (Q\,R_s)}{Q\,R_s} \bigg/ \sum_{s}f_s.
\qe
Figure~\ref{fig:lineprofiles} shows part of the calculated line profiles for some of the simulated materials in which the significant effect of the defect content on peak shape is clearly visible.

We now have a set of simulation-based samples of deformed material for which we know the exact defect content and have the corresponding diffraction line profiles. All that remains is to apply line profile analysis and compare the inferred defect content to the known, correct answer.

X-ray diffraction line profile analysis relies on the fact that the shapes of the diffraction peaks for a sample of crystalline material depend on the size of the crystallites and on the defects that they contain. The scattered intensity for a given diffraction peak (indexed by $hkl$) is a convolution of the effects of size and strain: $I_{hkl} = I_{hkl}^{\tt{size}} \ast I_{hkl}^{\tt{strain}}$. These two effects can be separated because they have a different dependence on diffraction order~\cite{Warren:1952aa}. The well known Williamson-Hall method \cite{Williamson:1953aa} provides a simple analysis by considering the full width at half maximum and integral breadth of each diffraction peak, but thereby reduces the change in shape of the peaks to two only parameters. Alternatively the Warren-Averbach method~\cite{Warren:1950aa,Warren:1959aa} retains full information about peak shape by considering the Fourier coefficients of the peaks $A_{hkl}(L) = A_{hkl}^{\tt{size}}(L)\, A_{hkl}^{\tt{strain}}(L)$, where $L$ denotes the Fourier variable. The strain broadening can be further decomposed into contributions from different types of defect We can write, for example, $A_{hkl}^{\tt{strain}}(L) = A_{hkl}^{\tt{disl}}(L)\, A_{hkl}^{\tt{SF}}(L)$,
for contributions from dislocations and stacking faults. This decomposition is possible because the Fourier coefficients for different defects vary in different ways from reflection to reflection ($hkl$-dependence) and with diffraction order.

A number of groups have implemented the Warren-Averbach method \cite{Delhez:1976wh,Berkum:1994ts,Balzar:2004tg} and we make use of the approach named \emph{convolutional multiple whole profile} (CMWP) analysis developed by Ungar and co-workers~\cite{Ungar:2001aa,Ungar:1999aa,Ungar:1984aa}. CMWP uses physically motivated profile functions for the Fourier coefficients $A_{hkl}(L)$ for different defect types, parameterised in terms of their density and character (e.g.\ burgers vector). The defect density is then inferred by optimising the values of the parameters so that they give a theoretical line profile that best matches the experimental profile. In the case of dislocations, Warren and Averbach \cite{Warren:1950aa} showed that the effect of strain gives rise to Fourier coefficients of the form
\eq\label{eq:wa14}
A_{hkl}^{\tt{disl}}(L) \approx \exp (-G_{hkl}^2 L^2\langle \varepsilon^2 \rangle_L /2),
\qe
where $G_{hkl}$ is the centre of the diffraction peak and $\langle \varepsilon^2 \rangle_L$ is the mean-square strain measured between pairs of points in the material separated by a distance $L$ in a direction perpendicular to the $(hkl)$ planes. This form for $A_{hkl}^{\tt{disl}}(L)$ also assumes that we consider only small values of $L$ and small strains. The derivation of \eqref{eq:wa14} is rather involved and can be found in Ref \cite{Warren:1950aa} in which \eqref{eq:wa14} appears as Equation 14.

Wilkens \cite{Wilkens:1970aa} derived a form for the mean square strain for a so-called restrictedly-random distribution of dislocations, in which dislocations are randomly distributed within ``cells'' of material,
\eq\label{eqn:msqstrain}
\langle \varepsilon^2 \rangle_L \approx
\frac{\rho C b^2}{4\pi}\ln\left( \frac{R_{\tt{e}}}{L} \right),
\qe
where $\rho$ is the dislocation density, $b$ is the length of the burgers vector and $R_{\tt{e}}$ is the size of the ``cell'' for the restrictedly-random distribution. $C$ is an average contrast factor, which quantifies the strength of the effect of a given dislocation type on the diffraction peak under study. The fundamental origins of this Krivoglaz-Wilkens strain function in ~\eqref{eqn:msqstrain} are detailed in Refs.~\cite{Groma:1997tk,Groma:2016uc,Zaiser:2001ur}. We investigate the interpretation of $R_{\tt{e}}$ further in our companion paper \cite{Race:ts}. The CMWP method is described in detail by Ungar et al.\ \cite{Ungar:2001aa} and Ribarik et al.\ \cite{Ribarik:2020aa}.

\begin{figure}\begin{center}
{\includegraphics[width=0.49\widefigwidth]{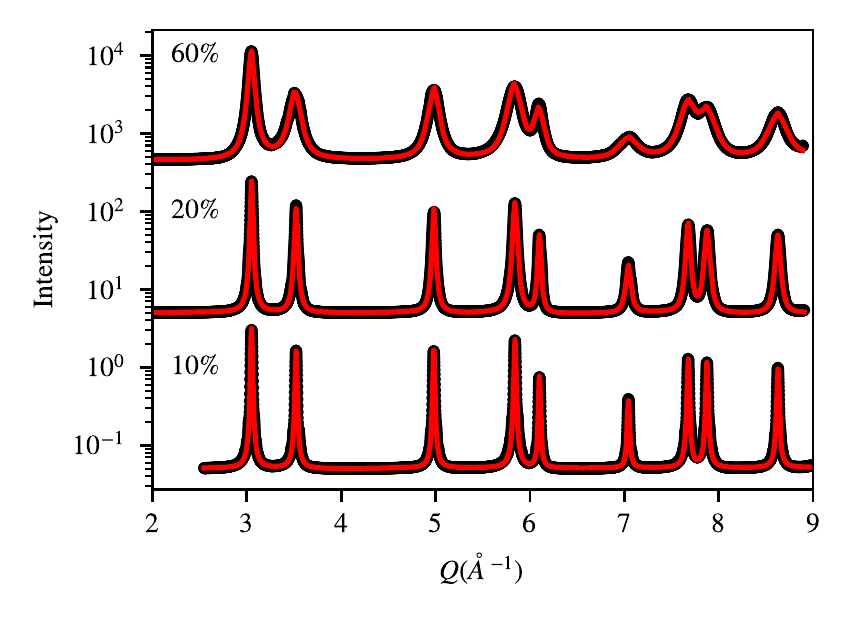}}\hfill
\caption{Comparison of original line profiles (thick, black lines) with CMWP fits (thin, red lines) for samples annealed for $0.5\,\tt{ns}$ at several strains. Profiles are offset vertically to aid readability. Note the log scale in intensity.}
\label{fig:cmwpfits}
\end{center}
\end{figure}

Figure~\ref{fig:cmwpfits} shows a comparison between the original diffraction profiles generated from the simulation cells and the theoretical profiles corresponding to the optimised values of the parameters in the model employed in the CMWP process. In all cases, for a variety of strains, these fits are very good, suggesting that there are no features peculiar to the profiles from the simulated material that will prevent successful analysis via the CMWP process.

\begin{figure}\begin{center}
\subfloat[]{\includegraphics[width=\figwidth]{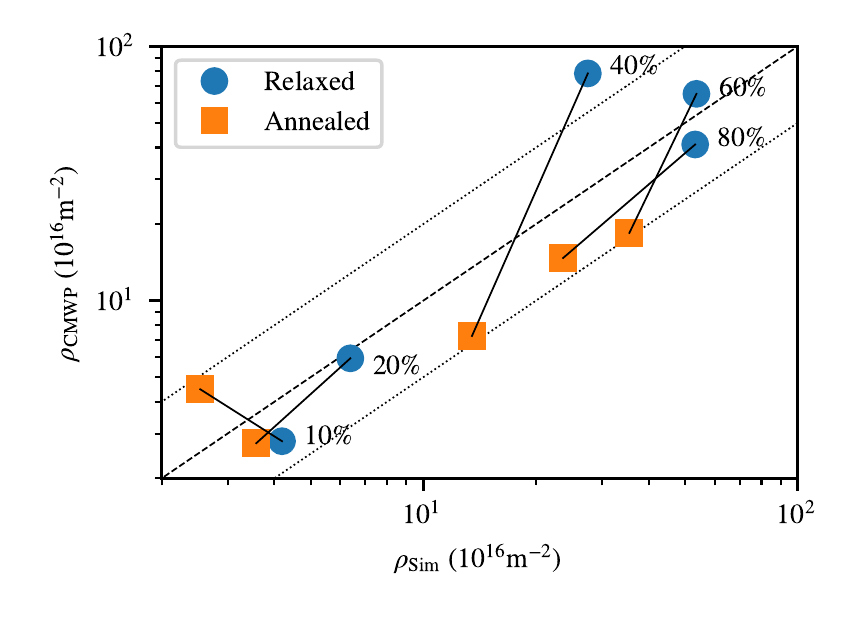}}\hfill
\subfloat[]{\includegraphics[width=\figwidth]{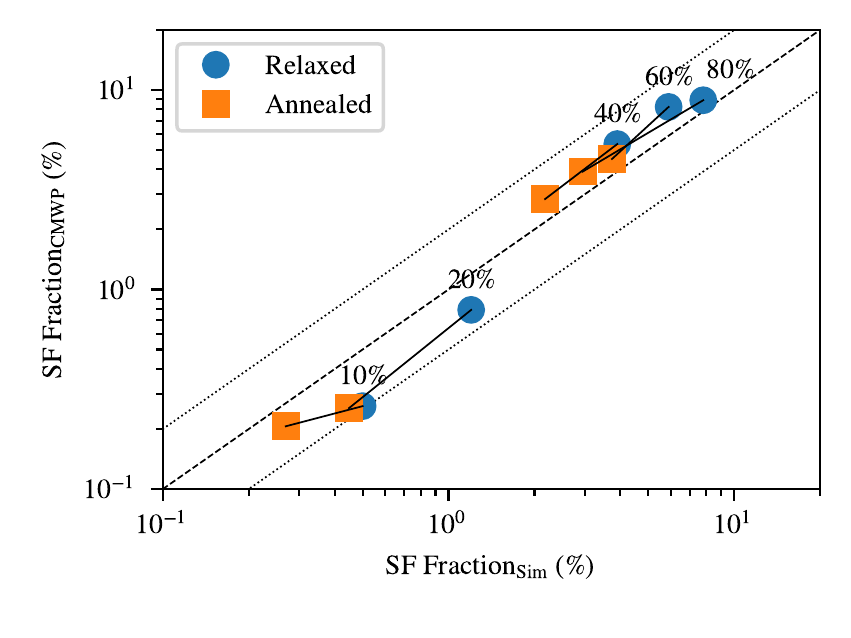}}
\caption{A comparison of (a) dislocation densities and (b) stacking fault fractions derived from line profile analysis with CMWP and the result of atom-by-atom analysis for the relaxed (blue circles) and $0.5\,\tt{ns}$ annealed (orange squares) cells. Solid black lines join relaxed and annealed points corresponding to the same initial strain to aid reading. The dashed black lines indicate perfect agreement ($X_{\tt{Sim}}=X_{\tt{CMWP}}$) between the atomistic analysis and CMWP and the dotted lines indicate a difference of a factor of 2.}
\label{fig:comparisonrhosf}
\end{center}
\end{figure}

We are now in a position to make a comparison between the defect content inferred via the CMWP method and the true values, calculated from the simulation cells. Figure~\ref{fig:comparisonrhosf}(a) shows the comparison for the dislocation densities for the relaxed and annealed material. Though quantitative agreement is not perfect, the true dislocation density varies over two orders of magnitude and CMWP does a good job of detecting this variation, generally being within a factor of 2 of the true value.

Figure~\ref{fig:comparisonrhosf}(b) shows a comparison between the CMWP stacking fault fraction and the fraction of atoms classified as having local hexagonal close packed coordination in a common neighbour analysis. We see excellent agreement across almost two orders of magnitude of stacking fault fraction.

By using atomistic simulations of cold-worked copper crystals to create material for which we can exactly quantify the defect content, we are able to directly test the performance of an important experimental method. We find that the CMWP approach to line profile analysis (an implementation of the Warren and Averbach approach) is able to simultaneously give an accurate measure of both dislocation and stacking fault content across several orders of magnitude of variation of defect density. Our results lend important new support to the use of line profile analysis in the examination of deformed crystalline material. Our companion paper \cite{Race:ts} contains further details of the methods and analysis and investigates other aspects of the line profile analysis, in particular probing the interpretation of the outer cut-off radius ($R_{\textrm{e}}$ in~\eqref{eqn:msqstrain}) in the physical model for dislocation strain fields used in the Warren and Averbach approach.

\begin{acknowledgments}
C.P.R. was funded by a University Research Fellowship of The Royal Society. T.U. was funded by the EPSRC Programme Grant (MIDAS: EP/S01702X/1). G.R. is grateful for the support of OTKA grant K124926 funded by the Hungarian National Research, Development and Innovation Office (NKFIH). Simulations were carried out on the University of Manchester's Computational Shared Facility. The core research data, along with computer scripts implementing our methods, are freely available for download \cite{Race:tb}.
\end{acknowledgments}


\bibliographystyle{unsrt}
\bibliography{references}
%

\end{document}